\documentclass[journal]{IEEEtran}
\usepackage[dvipsnames]{xcolor}
\usepackage{cite}
\usepackage{amsmath,amssymb,amsfonts}
\usepackage{mathtools}
\usepackage{lipsum}
\usepackage{algorithmic}
\usepackage{graphicx}
\usepackage{textcomp}
\usepackage{booktabs}
\usepackage{esint}

\ifCLASSINFOpdf
\else
\fi
\graphicspath{{Figures/}}

\begin{document}
%
\title{Equivalence of Polarizability and Circuit Models for Waveguide-Fed Metamaterial Elements}
%
%
%

\author{David R. Smith,~\IEEEmembership{Member,~IEEE,}
        ~Mohsen Sazegar,~\IEEEmembership{Member,~IEEE,} and Insang Yoo,~\IEEEmembership{Member,~IEEE}%
\thanks{D. R. Smith is with the Department
of Electrical and Computer Engineering and the Center for Metamaterials and Integrated Plasmonics, Duke University,
Durham, NC, 27708 USA. I. Yoo is with the School of Electrical and Electronic Engineering, Yonsei University, Seoul, Korea. He was previously with the Department of Electrical and Computer Engineering and the Center for Metamaterials and Integrated Plasmonics, Duke University, Durham, NC, 27708 USA. e-mail: insang.yoo@yonsei.ac.kr. M. Sazegar is with Kymeta Corp., Woodinville, WA, 98072, USA}
\thanks{Manuscript received June 10, 2024; revised June 10, 2024.}}

\maketitle

\begin{abstract}
A common variant of a metasurface antenna consists of an array of metamaterial elements coupled to a waveguide feed. The guided wave excites the metamaterial elements, coupling energy from the waveguide mode to radiation. Under appropriate conditions, each sub-wavelength metamaterial element can be modeled as a polarizable dipole, with the polarizability determined by an extraction procedure from the computed or measured waveguide scattering parameters. Here we establish the equivalence of this polarizability description of a metamaterial element with an equivalent circuit model, providing an additional tool for metasurface design that offers significant insight and a path towards efficiently modeling very large apertures. With this equivalence established, more complicated external circuits that include lumped elements and devices such as diodes and transistors can be integrated into the metamaterial element, which can then be transformed into an equivalent polarizability for modeling in the coupled dipole framework. We derive appropriate circuit models for several basic metamaterial elements, which provide direct relationships between the equivalent circuit parameters of an element and its effective polarizability. These expressions are confirmed using scattering parameters for several example structures obtained via full-wave simulations.
\end{abstract}


%
\IEEEpeerreviewmaketitle

\section{Introduction}
\begin{figure}[!b]
\centering
\includegraphics[width=2.75in]{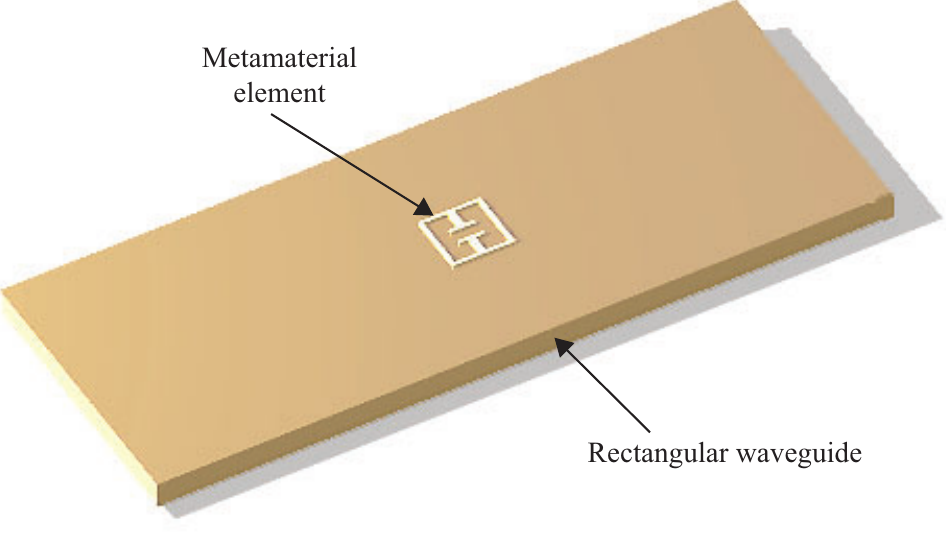}
\caption{A waveguide-fed metamaterial element in the top conducting surface of a rectangular waveguide.}
\label{fig:waveguide_fed_iris}
\end{figure}

%
%
%
%
\IEEEPARstart{M}{etasurfaces} have gained considerable traction as a viable antenna paradigm \cite{holloway2012overview, wang2020metantenna, maci2011metasurfing, minatti2014modulated, quevedo2019roadmap, glybovski2016metasurfaces, li2018metasurfaces}, particularly for large apertures where their inherent passive architecture provides considerable benefit. One metasurface configuration that lends itself especially well to the format of aperture antennas is the \emph{waveguide-fed metasurface}, in which a waveguide is used to excite a collection of metamaterial elements located on one of the conducting walls of the waveguide \cite{smith2017analysis, li2020widebandmetasurface, johnson2014discrete}. The waveguide-fed metasurface antenna can be considered a variant of a travelling-wave or leaky-wave antenna, but the inherent control over the phase and magnitude of the fields radiated from each metamaterial iris---achieved through the use of resonant elements---distinguishes the waveguide-fed metasurface antenna from either of these prior constructs. The commonality among these antennas is a waveguide mode within an enclosed or semi-enclosed feed structure having a propagation constant $\beta$, where

\begin{equation}
    \beta = \frac{\omega n_{wg}}{c},
\end{equation}

\noindent and $n_{wg}$ is the waveguide index. If a slot, iris or metamaterial element (here all treated as synonymous and illustrated in Fig. \ref{fig:waveguide_fed_iris}) is introduced at some point along the waveguide, then part of the energy from the waveguide mode will couple out of the structure via the element while part will be scattered back into the waveguide. Note that an array of such elements differs from a phased array or active electronically steered antenna (AESA) by the absence of powered components, such as phase shifters or amplifiers; energy is injected into the antenna solely via the feed mode. Ideally, only a small fraction of the incident energy is scattered and radiated by a given metamaterial element, such that the overall aperture efficiency of the composite aperture is high. Throughout the following discussion, we consider a metamaterial element as being complementary, with air and metal swapped compared to a volumetric metamaterial element \cite{falcone2004babinet}. The purpose of the metamaterial element is to provide a radiating node whose phase and amplitude can be controlled (though not independently) through its resonance by adjusting its geometry or electrical configuration.

Each metamaterial element is assumed to be much smaller than either the free-space or waveguide wavelength, spaced apart by distances equal to or smaller than half the free-space wavelength. Under this assumption the metamaterial element can often be approximated as a polarizable dipole, with polarizability tensors $\bar{\bar{\alpha_e}}$ and $\bar{\bar{\alpha_m}}$. This description of the element can be built into a multiscale modeling tool for large and complex apertures, referred to here as the coupled dipole model \cite{bowen2012using,pulido2017polarizability}. The advantage of the coupled dipole model is that numerical simulations need only be performed on a single metamaterial element and a small section of waveguide, from which the effective polarizability components are extracted. Subsequently, radiated far-fields and other quantities from apertures comprising many (hundreds or thousands) of metamaterial elements can be determined efficiently by summing over the individual dipole contributions \cite{yoo2019analytic, yoo2020full}. 


Of particular relevance to this work is the development of dynamically reconfigurable metasurface aperture antennas \cite{boyarsky2021electronically,yoo2023experimental}, in which a tuning component---such as a varactor or a diode---is incorporated into each metamaterial element providing dynamic tuning. We take as an example the varactor, which introduces a voltage-controlled capacitance to the metamaterial element. While such lumped elements can be included in full-wave simulations of metamaterial elements, it is not as easy to gain a quick, intuitive understanding of the behavior of the composite element, since the lumped element is described by a circuit model while the metamaterial element is described by a polarizability. The design of the composite element then relies on numerous numerical simulations in which, for example, a varactor capacitance is varied over its tuning range, with the polarizability of the composite then interpolated from these simulations. As the circuitry added to the metamaterial element becomes more complicated---adding diodes, transistors, matching networks, switches, or other components---the reliance on full-wave modeling increases. There is thus a need for a means of converting the polarizability description to a circuit description, and the reverse. If an extracted polarizability can be converted to a circuit model, the effect of additional lumped circuit components incorporated into the element becomes easy to understand and more advanced design can be facilitated. An element design, including all integrated circuit components, can then be converted back to the polarizability model for quick and efficient modeling of the composite aperture---potentially obviating hundreds of numerical calculations and associated interpolation.

In the present work, we develop a method of transforming the extracted polarizability of a waveguide-fed metamaterial element to an equivalent circuit model. Since a metamaterial element can be viewed as a collection of interacting slots, prior work on equivalent circuit models of slot antennas is instructive \cite{stevenson1948waveguideslots, oliner1957impedance, pozar1986microstripfedslot, eshrah2007slotcircuitmodel}. In an early work on the topic, Oliner applied a variational approach to derive an equivalent circuit model for resonant and non-resonant rectangular slots inserted into a waveguide feed, for various positions and orientations \cite{oliner1957impedance}. Oliner \cite{oliner1957impedance} and Stevenson \cite{stevenson1948waveguideslots} found that a rectangular waveguide with a transverse rectangular slot in the broad face can be modeled as a transmission line with a series impedance; they further determined analytical expressions for the susceptance and conductance of this equivalent impedance. Numerous investigations have also demonstrated similar results for microstrip-fed slots in ground planes, where again the slot can be modeled as a series impedance \cite{pozar1986microstripfedslot}. For waveguides with thick walls, the electrical model includes a transformer element to account for the coupling from waveguide to free space. Even in the absence of analytical expressions for the relevant circuit parameters, the use of a circuit model with full-wave simulations can provide a useful and intuitive understanding of the behavior of a slot antenna or coupling element, as illustrated by Eshrah \emph{et al.} \cite{eshrah2007slotcircuitmodel}. The latter approach exchanges the mathematical complexity of the analytical models in favor of full-wave simulations; such simulations are easily performed over small computational domains, providing an efficient means of determining the circuit model of a slot. Because the waveguide-fed slot antenna is so well-characterized, we use this configuration to develop the fundamental connection between polarizability and circuit models.

The effective polarizability of a metamaterial element can be obtained from the computed or measured scattering (S-) parameters at the input and output ports of a waveguide. Likewise, an element with known polarizability will produce a unique set of S-parameters. Using the relationship between the S-parameters and the transfer (or ABCD) matrix, a circuit model can be inferred for the element. Although explicit analytical expressions are not derived here for the element circuit parameters in terms of the geometrical features, certain fundamental aspects of the element become apparent through this analysis---particularly the role that radiation damping plays and its impact on the analytical equations for both the static and dynamic polarizabilities. The form of the dynamic polarizability, for example, is used to identify a simple connection between the equivalent impedance and polarizability descriptions of the metamaterial element.

Building on the mathematical framework presented in \cite{collin1990fieldCh7}, a computational polarizability extraction strategy was developed previously for a metamaterial element embedded in a rectangular waveguide \cite{pulido2017polarizability}. Rather than repeating the details of this approach, we instead summarize the salient features in the introduction to Section \ref{sec:polarizability_extraction}, highlighting the key equations needed for the subsequent analysis. In the appendix, we also provide an alternative, approximate derivation that motivates the key polarizability extraction equations. For this work, we adopt the rectangular waveguide feed, which can be applied as well to printed circuit board (PCB) metasurface antenna variants using the substrate integrated waveguide (SIW) format. Other waveguide feeds can be analyzed in the same manner as presented here using the Green's function appropriate for the particular guide. We note that much of the development presented here parallels that of \cite{collin1990fieldCh7}, where the equivalence of polarizability and circuit models is established in the context of waveguides coupled by small apertures. Our aim here is to extend these analytical methods into a more general computational tool for advanced metasurface apertures. 

\section{Polarizability Extraction}
\label{sec:polarizability_extraction}

\begin{figure}[!b]
	\centering
	\includegraphics[width=2.25in]{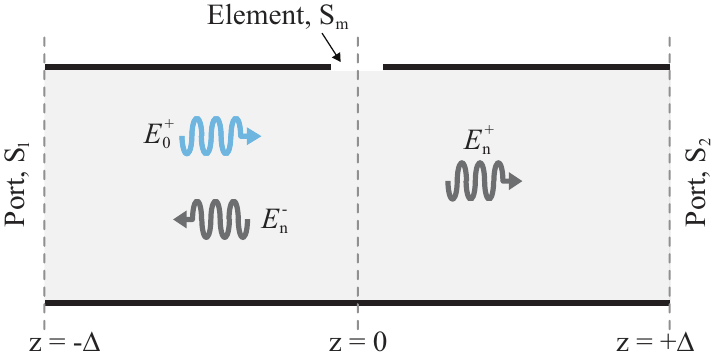}
	\caption{\label{fig:mode_description}Incident and scattered modes in a rectangular waveguide.}
\end{figure}

Polarizability extraction refers to the numerical procedure by which effective polarizabilities are obtained for a metamaterial element. As described above, the polarizability model presumes that the element can be described as a point dipole, which must typically be first confirmed with full-wave numerical simulations. For the element to be described as a point dipole---or a collection of point dipoles, its dimensions should be sufficiently subwavelength that its fields are well-described by the appropriate Green's functions both within and without the waveguide. With the point dipole assumption in place, the effective polarizabilities to be ascribed to an element can be inferred from the transmitted and reflected waves within the waveguide feeding the metamaterial element. The configuration for polarizability extraction is shown schematically in Fig. \ref{fig:mode_description}, which illustrates the cross-section of a waveguide with the incident electric field and scattered $n$th mode shown. For this analysis, we assume a rectangular waveguide operating at frequencies above the cutoff frequency of the lowest TE mode and below the cutoff frequency of the first higher order mode. In this situation there is only one mode that propagates---the $\textrm{TE}_{10}$ mode---which has the form

\begin{equation}
\begin{aligned}
E_y^{\pm}(x,y,z) &= -\frac{j \omega \mu \pi}{k_c^2 a}C_{10}\sin{\frac{ \pi}{a}x}e^{\mp j \beta z}, \\
H_x^{\pm}(x,y,z) &= \frac{j \beta \pi}{k_c^2 a}C_{10}\sin{\frac{\pi}{a}x}e^{\mp j \beta z}, \\
H_z^{\pm}(x,y,z) &= C_{10}\cos{\frac{\pi}{a}x}e^{\mp j \beta z},
\end{aligned}    
\end{equation}

\noindent where

\begin{equation}
    k_c^2 = \frac{\omega^2}{c^2}-\beta^2=\left( \frac{m\pi}{a}\right)^2+\left( \frac{n\pi}{b}\right)^2,
\end{equation}

\noindent and $E_y^+$ indicates a forward propagating wave while $E_y^-$ indicates a backward propagating wave.

A metamaterial element can be viewed as a perturbation to the waveguide that will generally excite all modes, so that the field produced by scattering from the element must rigorously be expressed a sum over all of the possible modes. Since all other modes aside from the $\textrm{TE}_{10}$ mode are beyond cutoff, they will decay away from the element exponentially, leading to reactive fields that contribute to an effective lumped inductance or capacitance. 

In the dipole framework, it is the effective dipole moments $\vec{p}$ and $\vec{m}$ of the metamaterial element that determine both the scattered and radiated fields. The effective dipole moments associated with the metamaterial can be related to the incident fields by the polarizabilities according to

\begin{equation}
\begin{aligned}
\vec{p} &= \epsilon \bar{\bar{\alpha}}_e \vec{E}_0^{+}, \\
\vec{m} &= \bar{\bar{\alpha}}_m \vec{H}_0^{+}.
\end{aligned}
\end{equation}

\noindent Given that we consider only TE modes, the only possible excitation of a dipole within the surface of the waveguide is via either the electric field out-of-plane with the aperture or the magentic field component in the plane of the aperture. Thus, to describe the scattering of the metamaterial aperture, we need only consider the out-of-plane electric polarizability and the in-plane magnetic polarizability tensor components. Furthermore, if the metamaterial aperture is placed along the symmetry point on the $x$-axis, there will be no field component exciting the element along the propagation direction, so that the polarizability component along that direction need not be considered \cite{pulido2017polarizability}. The result is that only $\alpha_{ey}$ and $\alpha_{mx}$ are of concern and need to be calculated.

To relate the two polarizability components to the transmitted and reflected fields in the waveguide, we first apply the normalization condition 

\begin{equation}
    \label{eq:mode_norm}
    \int_S \vec{E}_t \times \vec{H}_t \cdot d \vec{s} = \int_S E_y H_x ds = \frac{1}{Z_0},
\end{equation}

\noindent so that the mode coefficient is found as

\begin{equation}
    C_{10}^2 = \frac{1}{Z_0}\frac{2}{a b} \frac{k_c^2}{\omega \mu \beta}.
\end{equation}

\noindent The relevant electric and magnetic fields are then

\begin{equation}
\begin{aligned}
\label{eq:incident_fields}
E_{0y} &= -j \sqrt{\frac{2}{a b}}\sin{\frac{\pi x}{a}}e^{-j \beta z}, \\
H_{0x} &= -j \sqrt{\frac{2}{a b Z_0^2}}\sin{\frac{\pi x}{a}}e^{-j \beta z},
\end{aligned}
\end{equation}

\noindent and the fields at the position of the metamaterial element are

\begin{equation}
\begin{aligned}
E_{0y}^2 &= -\frac{2}{a b}, \\
H_{0x}^2 &= -\frac{2 }{a b Z_0^2}.
\end{aligned}
\end{equation}

\noindent To this point, our discussion has been in terms of the physical waveguide modes, with all quantities expressed in terms of the waveguide electric and magnetic fields. We can alternatively describe the waveguide as a transmission line with an effective voltage $V$, current $I$, and line impedance $Z_L$. In the following, we adopt the convention in \cite{pozar2009microwave} to equate the two descriptions, with the choice of $Z_L=Z_0$ for the effective line impedance. 

The polarizabilities can be written in terms of the scattering parameters as \cite{pulido2017polarizability}

\begin{equation}
\label{eq:polarizabilities_vs_s_parameters}
\begin{aligned}
    \alpha_{ey} &=\frac{j a b \beta}{2 k^2}\left(S_{21}+S_{11}-1\right), \\
    \alpha_{mx} &=\frac{j a b}{2 \beta}\left(S_{21}-S_{11}-1\right).
\end{aligned}
\end{equation}

These equations, also derived here in the appendix, form the basis of a procedure for extracting the equivalent polarizability of a metasurface element from the scattering parameters within the waveguide feed, or \emph{polarizability extraction}. For reference later, the scattering parameters can be expressed in terms of the dipole polarizabilities as

\begin{equation}
\begin{aligned}
S_{21} &= \left(1-j\frac{k^2 \alpha_{ey}}{a b \beta}-j \frac{\beta \alpha_{mx}}{a b}\right), \\
S_{11} &= \left(-j\frac{k^2 \alpha_{ey}}{a b \beta}+j \frac{\beta \alpha_{mx}}{a b}\right).
\end{aligned}
\end{equation}


The equations relating the polarizabilties in terms of the scattering parameters are sufficient as a means of characterizing a waveguide-fed metamaterial element for cases where the dipole approximation provides a suitable description of the element. The extent to which this approximation is valid relies in turn on the validity of numerous assumptions in the coupled mode theory presented in \cite{pulido2017polarizability} as well as in \cite{collin1990fieldCh7}. These are not inconsequential assumptions, such that a waveguide-fed metasurface aperture should be designed from the beginning with these constraints in mind.

\subsection{The Impact of Radiation Damping}
We assume initially that the metals are perfect conductors, that there are no other lossy materials within the waveguide structure, and ignore radiation to free space. We initially proceed with this assumption, noting that
 (\ref{eq:polarizabilities_vs_s_parameters}) provides an exact relationship between the scattering parameters and the effective polarizabilities that describe the metamaterial element as a dipole.

In the absence of any apparent losses, one might first assume the dipole polarizabilities are purely real quantities. However, energy scattered by the element into the waveguide can be considered a loss mechanism, necessitating the introduction of an imaginary part to each of the polarizabilities---the radiation damping terms. We first write (\ref{eq:polarizabilities_vs_s_parameters}) in a dimensionless form, introducing the dimensionless polarizabilities $A_e$ and $A_m$,

\begin{equation}
\begin{aligned}
    -j\frac{2k^2 \alpha_{ey}}{a b \beta} &= S_{21}+S_{11}-1=-j A_e, \\
    -j\frac{2 \beta \alpha_{mx}}{a b} &= S_{21}-S_{11}-1=-j A_m,
\end{aligned}
\end{equation}

\noindent where

\begin{equation}
\begin{aligned}
A_m &= \frac{2 \beta \alpha_m}{a b}, \\
A_e &= \frac{k^2}{\beta}\frac{2 \alpha_{e}}{a b}.
\label{eq:normalized_polarizabilities}
\end{aligned}
\end{equation}

\noindent The dimensionless polarizabilities relate very simply to the scattering parameters and provide a more intuitive indication of the scattering strength of the dipole. 

The metamaterial elements of interest here are typically dominated by their magnetic polarizability, so we assume $\alpha_{mx}$ is real and set $\alpha_{ey}=0$ for the initial discussion. Since the incident, transmitted and reflected waves are all determined via (\ref{eq:polarizabilities_vs_s_parameters}) (assuming the incident power is normalized to unity), energy conservation then requires $|S_{21}|^2+|S_{11}|^2-1=0$. Calculating this quantity explicitly, we find

\begin{equation}
    |S_{21}|^2+|S_{11}|^2 - 1=|A_m|^2/2 = 2 \left(\frac{\beta \alpha_{mx}}{a b}\right)^2,
\end{equation}

\noindent which clearly does not conserve energy. The resolution to this apparent paradox is to introduce an imaginary term into the expression for the polarizability to account for the radiative loss, essentially transforming the \emph{static} polarizability to the \emph{dynamic} polarizability. While there are two loss mechanisms---radiation into the waveguide and into free space---we consider first just the contribution of radiation into the waveguide. Althought the corrected polarizability---termed the \emph{dynamic polarizability}---can be computed simply from knowledge of the waveguide Green's function, we provide an alternative derivation based on energy conservation that provides some useful insights. Note that this discussion also resembles that contained in \cite{collin1990fieldCh7}.

With $A_e=0$, 

\begin{equation}
\begin{aligned}
S_{21} &= 1-j\frac{A_m}{2}, \\
S_{11} &= j\frac{A_m}{2}.
\label{eq:spars_magnetic_dipole}
\end{aligned}
\end{equation}

\noindent Squaring each of the above terms yields

\begin{equation}
\begin{aligned}
|S_{21}|^2 &= 1+\frac{|A_m|^2}{4}+\textrm{Im}\{ A_m \}, \\
|S_{11}|^2 &= \frac{|A_m|^2}{4}.
\end{aligned}
\end{equation}

\noindent Insisting that energy conservation holds, or that $1-|S_{21}|^2-|S_{11}|^2=0$, a condition is found immediately on the dimensionless polarizability:

\begin{equation}
    \textrm{Im}\{ A_m \} = -\frac{|A_m|^2}{2}.
\end{equation}

\noindent This equation shows that the scattering of the incident wave into the waveguide must be balanced by an imaginary term in the polarizability that accounts for the power loss. Since we began by assuming the polarizability to be purely real, we continue this assumption and treat the imaginary part as being an additional contribution that accounts for the radiation damping within the waveguide. For this reason in the following we drop the absolute value signs on our initial polarizability value. In terms of the polarizability and waveguide parameters,

\begin{equation}
    \textrm{Im}\{\alpha_m\}=-\frac{\beta}{a b} \alpha_m^2.
    \label{eq:consistency_requirement}
\end{equation}

\noindent The dynamic polarizability thus necessarily has an imaginary component. However, one more correction to the polarizability is needed since the new imaginary term in the polarizability constitutes power dissipated by the dipole that has not yet been fully accounted for. The magnitude of this power is easily found as

\begin{equation}
    P_{\textrm{abs}}=\frac{\omega}{2}\textrm{Im}\{\vec{m}\cdot \vec{B}\}=\frac{\mu \omega}{2}\textrm{Im}\{\alpha_m\} |\vec{H}_0^{+}|^{2}.
\end{equation}

\noindent Using (\ref{eq:consistency_requirement}) along with the expression for the incident magnetic field (\ref{eq:incident_fields}), we obtain

\begin{equation}
\begin{aligned}
    P_{\textrm{abs}} &= \frac{\mu \omega}{2}
    \Big( \frac{\beta}{ab}\alpha_m^2\Big)
    \Big( \frac{2}{a b Z_0^2}\Big) \\
    &=\frac{A_m^2}{4 Z_0}.
\end{aligned}
\end{equation}


\noindent This quantity represents the actual power absorbed by the dipole, assuming the field normalization of (\ref{eq:mode_norm}). A quantity of greater relevance here would be the normalized power absorbed, which can be found by dividing the above quantity by $1/Z_0$, or

\begin{equation}
    \bar{P}_{\textrm{abs}}=\frac{A_m^2}{4}.
\end{equation}

\noindent Although the incident power was assumed fixed, the total scattered and absorbed power have been increased by the amount of power absorbed by the dipole. The total power should thus be corrected to $\bar{P}_{\textrm{inc}}=1+A_m^2/4$, in which case the correct ratio of absorbed to incident power is

\begin{equation}
    \bar{P}_{\textrm{abs}}=\frac{\frac{\mu \omega}{2}\textrm{Im}\{\alpha_m\}  |H_{\textrm{inc}}|^2}{1+A_m^2/4}.
\end{equation}

\noindent At this point, we could interpret the above equation as indicating the field at the dipole has been reduced by the factor $1+A_m^2/4$, such that energy balance is restored. Alternatively, we can apply the correction directly to the polarizability, assuming the incident power unchanged. Although the correction is just applied to the imaginary part of the polarizability in the above equation, it must be applied equally to the real part as well. Applying the correction to the dimensionless polarizability, we obtain

\begin{equation}
    \label{eq:dynamic_normalized_polarizability}
   \Tilde{A}_m \rightarrow \frac{A_m-jA_m^2/2}{1+A_m^2/4}=\frac{A_m}{1+jA_m/2}.
\end{equation}

\noindent or, for the polarizability,

\begin{equation}
    \Tilde{\alpha}_{mx} \rightarrow \frac{\alpha_{mx}}{1+j \frac{\alpha_{mx} \beta}{a b}}.
    \label{eq:sipe_polarizability}
\end{equation}

\noindent This final form for the polarizability is identical to the well-known Sipe polarizability for a radiating dipole and is consistent with prior analysis on small apertures in waveguides \cite{collin1990fieldCh7}.

In addition to radiating into the waveguide, a metamaterial element also radiates power to free space. The energy conservation equation must thus be modified as

\begin{equation}
    1-|S_{21}|^2-|S_{11}|^2-\bar{P}_{rad} = 0.
\end{equation}

\noindent Here, $\bar{P}_{rad}$ is the ratio of power radiated to that of the incident power. Using the expressions derived above, we have

\begin{equation}
    \textrm{Im} \{ A_m \} = -\frac{A_m^2}{2}-\bar{P}_{rad}.
\end{equation}

\noindent Just as with scattering within the waveguide, power radiated to free space must be accounted for by a second imaginary term in the polarizability. An analytic expression for the radiation damping term corresponding to a finite-width waveguide is not easily found; instead, we apply the Green's function for a dipole radiating into a half space above a conducting plane. While not the specific boundary condition of the waveguide structure, the free-space radiative damping term is well-known and can be introduced additively as follows 

\begin{equation}
    \Tilde{\alpha}_{mx} \rightarrow \frac{\alpha_{mx}}{1+j \alpha_{mx}\Big(\frac{ \beta}{a b} + \frac{k^3}{3\pi} \Big)}.
\end{equation}

\noindent In general, the ratio of the power radiated to free space relative to that of the incident power can be found from the expression

\begin{equation}
    \label{eq:radiated_power}
    \bar{P}_{rad}=-\frac{A_m^2}{2}-\textrm{Im} \{ A_m \}.
\end{equation}

\noindent This result is useful, since it does not rely on the exact analytical form of the free-space Green's function term. Rather, the radiated power (assuming no resistive losses in the element or guide) can be found by measuring or computing the normalized polarizabilities extracted from the waveguide scattering parameters.

When both polarizabilities are present, the above equations can be generalized as

\begin{equation}
    \textrm{Im} \{ A_m+A_e \} = -\frac{A_m^2+A_e^2}{2}-\bar{P}_{rad}.
\end{equation}

\section{Equivalent Circuit Model} \label{sec:circuit model}

Having related the scattering parameters to the equivalent polarizabilities associated with metamaterial apertures, it is useful to next relate the S-parameters to the transfer---or \emph{ABCD}---matrix. The reason for doing so allows for simple formulas for cascading multiple elements (assuming no other interactions other than scattering). However, it is also possible from observation of the transfer matrix elements to infer equivalent circuit models for the metamaterial apertures. Having the connection between a circuit model and its equivalent polarizability provides a powerful tool for metasurface design.

To clarify the analysis, we consider three cases: A magnetic dipole; an electric dipole; and the combination of a magnetic and electric dipole. By considering first single dipole components, the emergence of the impedance model is more easily understood.

\subsection{ABCD Matrix: Magnetic Dipole} \label{ABCD Matrix: Magnetic Dipole}

The scattering parameters for the magnetic dipole were presented in the previous section, (\ref{eq:spars_magnetic_dipole}). From these parameters it is straightforward to compute the transfer matrix---or \emph{ABCD} matrix---via the transformation \cite{pozar2009microwave}

\begin{equation}
    \begin{aligned}
        T_{11} &=\frac{(1+S_{11})(1-S_{22})+S_{21}S_{12}}{2 S_{21}}, \\
        T_{12} &=Z_0\frac{(1+S_{11})(1+S_{22})-S_{21}S_{12}}{2 S_{21}}, \\
        T_{21} &=\frac{1}{Z_0}\frac{(1-S_{11})(1-S_{22})-S_{21}S_{12}}{2 S_{21}}, \\
        T_{22} &=\frac{(1-S_{11})(1+S_{22})+S_{21}S_{12}}{2 S_{21}}.
    \end{aligned}
\end{equation}

\noindent Using (\ref{eq:spars_magnetic_dipole}), we find $T_{11} = T_{22} = 1$, $T_{21}=0$ and

\begin{equation}
    T_{12} = Z_0 \frac{2 j A_m/2}{1-j A_m/2}.
\end{equation}

\noindent The form of the ABCD matrix, with the diagonal elements equal to unity and $T_{21}=0$ implies a circuit model consisting of a series impedance \cite{pozar2009microwave}, allowing us to equate $T_{12}=Z$ with the resulting transfer matrix

\begin{equation}
    \bar{\bar{T}}=\begin{pmatrix}
    1 & Z  \\
    0 & 1  
    \end{pmatrix},
\end{equation}

\noindent where

\begin{equation}
    Z=Z_0 \frac{j A_m}{1-j A_m/2}.
    \label{eq:impedance}
\end{equation}

\begin{figure}[!b]
	\centering
	\includegraphics[width=1.85in]{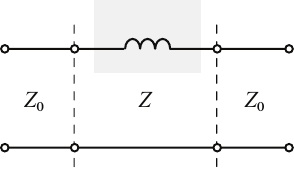}
	\caption{\label{fig:series_impedance_network}Equivalent circuit for the case of a magnetic polarizability.}
\end{figure}

\noindent The equivalent circuit model is shown in Fig. \ref{fig:series_impedance_network}. Since we have placed no restrictions on $A_m$ here, we may assume it to be complex. However, multiplying the top and bottom of (\ref{eq:impedance}) by $A_m^*$ we find

\begin{equation}
    Z=Z_0 \frac{j |A_m|^2}{A_m^*-j |A_m|^2/2}=jZ_0\frac{|A_m|^2}{\textrm{Re}\{A_m\}}.
    \label{eq:impedance_series}
\end{equation}

\noindent If there is no other radiative mechanism, then for consistency the imaginary part of the polarizability must compensate for the power radiated in the waveguide. The impedance is purely imaginary and is inductive. This result is consistent with irises and slots in waveguides, which can be modeled as inductive loads. If a radiative component exists, then, the equivalent circuit for the magnetic polarizability will be an equivalent inductance in parallel with a radiation resistance.

The equation for the impedance can be inverted, allowing the normalized polarizability to be found in terms of the impedance as

\begin{equation}
    A_m = \frac{-j 2 Z}{2 Z_0+Z}.
    \label{eq:norm_mag_pol_vs_z}
\end{equation}

\noindent For the non-resonant, non-lossy magnetic dipole, we expect the equivalent circuit model consists of an inductor so that the impedance is $Z=j\omega L$. Using this expression for $Z$ and $Z_0=\mu \omega/\beta$,

\begin{equation}
    \frac{A_m}{2} =  \frac{\frac{\beta L}{2 \mu}}{1+j\frac{\beta L}{2\mu}},
\end{equation}

\noindent or

\begin{equation}
    \alpha_m =  \frac{\frac{ab L}{2 \mu}}{1+j\frac{\beta L}{2\mu}}=\frac{\frac{ab L}{2 \mu}}{1+j\left(\frac{a b L}{2\mu}\right)\left(\frac{\beta}{a b}\right)}.
    \label{eq:magnetic_polarizability_circuit_equiv}
\end{equation}

\noindent Note that although we have assumed a purely reactive impedance, the effective polarizability has exactly the form of (\ref{eq:sipe_polarizability}), leading us to conclude the static polarizability is

\begin{equation}
    \alpha_{m0} = \frac{a b L}{2 \mu}.
    \label{eq:static_polarizability}
\end{equation}

\noindent This final equation---similar to that found in \cite{collin1990fieldCh7}---provides a simple relationship between the polarizability and circuit models. Once the polarizability has been extracted, within the same level of approximation one can ascribe an effective lumped inductance as an alternative description. Note that the strength of the polarizability relates not only to the effective inductance, but also to the waveguide properties, the latter providing an indication of coupling to the element.

In general, a slot or any iris always possesses a resonance at some higher frequency, so that its polarizability is always dispersive to some extent. To better model a slot, as will be shown below, some capacitance should be included within the slot. Nevertheless, (\ref{eq:static_polarizability}) provides a useful conceptual framework for the circuit model of a non-resonant slot, underlying the fact that a magnetic polarizability is predominantly inductive in character.

\subsubsection{The Impact of Radiation Damping} \label{Radiation Damping}

The effective impedance for a magnetic polarizability is inherently complex due to the presence of radiation damping both into the waveguide and into free space. The presence of the radiation damping terms lead to an inherent dispersion to both the real and imaginary parts of the impedance, as implied by (\ref{eq:impedance}). Then, $\alpha_m$ in (\ref{eq:magnetic_polarizability_circuit_equiv}) can be written as

\begin{equation}
    \alpha_m = \frac{\alpha_{m0}}{1+j\alpha_{m0} (\frac{\beta}{a b} + \frac{k^3}{3\pi})},
\end{equation}

\noindent where we have included the free space radiation damping term. Rewriting (\ref{eq:impedance}) as an admittance yields

\begin{equation}
    Y = \frac{1}{Z} = \frac{1}{Z_0}\Big( -\frac{j}{ A_m} - \frac{1}{2}\Big)
\end{equation}

\noindent or

\begin{equation}
    Y = \frac{1}{2 Z_0}\Big( -\frac{j}{\frac{\beta}{a b} \alpha_m} -1\Big)
\end{equation}

\noindent from which we conclude

\begin{equation}
    Y  = \Big( \frac{1}{j \omega L} +\sqrt{\frac{\epsilon}{\mu}}\frac{a b k^2}{6 \pi}
    \Big)
\end{equation}

\begin{figure}[!b]
	\centering
	\includegraphics[width=1.85in]{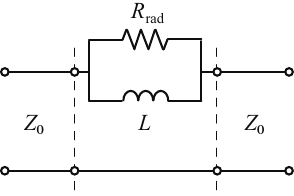}
	\caption{\label{fig:series_impedance_radiation_damping}Equivalent circuit for the case of a magnetic polarizability when radiation damping is included.}
\end{figure}

The result of this analysis shows that radiation damping presents itself as a radiation resistance in parallel with the effective inductance, leading to the circuit model shown in Fig. \ref{fig:series_impedance_radiation_damping}. The radiation resistance is seen to have the value

\begin{equation}
    R_{rad}=\sqrt{\frac{\mu}{\epsilon}} \frac{6\pi }{a b k^2}
    \label{eq:radiation_resistance_formula}
\end{equation}

\noindent Interestingly, this expression is nearly identical with that derived by Stevenson (see (44) in \cite{stevenson1948waveguideslots}) although arrived at via an entirely different approach. The agreement stems from the fundamental manner in which energy conservation is satisfied for the lossless waveguide and the radiation damping terms, which leads to an expression largely independent of the particular slot geometry.

\subsubsection{Resonant Magnetic Polarizability} \label{ABCD:Resonant Magnetic Polarizability}

If we now allow for a more general series impedance, beyond just a single lumped inductor, we can generalize (\ref{eq:norm_mag_pol_vs_z}) as follows. With $Z_0 = \mu \omega/\beta$, we have

\begin{equation}
    \frac{A_m}{2} = \frac{-jZ}{2\frac{\mu \omega}{\beta}+Z},
\end{equation}

\noindent or

\begin{equation}
    \frac{A_m}{2} = \frac{-j \frac{\beta Z}{2 \mu \omega}}{1+\frac{\beta Z}{2 \mu \omega}}.
\end{equation}

\noindent The polarizability is then

\begin{equation}
    \alpha_m = \frac{-j \frac{a b Z}{2 \mu \omega}}{1+\frac{a b Z}{2 \mu \omega}\left( \frac{\beta}{a b} \right)}.
\end{equation}

\noindent The static polarizability thus relates to the equivalent series impedance as

\begin{equation}
    \alpha_{m0} = -j\frac{a b Z}{2 \mu \omega}.
    \label{eq:generalized_polarizability}
\end{equation}

\begin{figure}[!b]
	\centering
	\includegraphics[width=1.85in]{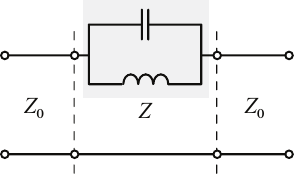}
	\caption{\label{fig:series_impedance_resonant_network}Equivalent circuit for the case of a simple resonant polarizability.}
\end{figure}

Consider first the case of a resonator formed by a parallel combination of inductance and capacitance, as shown in Fig. \ref{fig:series_impedance_resonant_network}. Such a model would describe, for example, a rectangular slot where its inherent capacitance is included. If the frequencies of interest are significantly below the resonance frequency, then the capacitance will serve as a perturbation on the otherwise inductive polarizability. For more complex elements, such as the cELC, a simple LC model is unlikely to apply, but may yet provide a useful conceptual model of the element for certain situations. 

For the parallel inductor/capacitor case, the impedance is 

\begin{equation}
    Z^{-1}=\frac{1}{j \omega L}+j \omega C.
\end{equation}

\noindent so that,

\begin{equation}
    \label{eq:resonant_polarizability}
    \alpha_m = -j \frac{a b}{2 \mu \omega}\left[ \frac{1}{\frac{1}{j \omega L}+j \omega C} \right]=\frac{a b L \omega_0^2}{2\mu}\frac{1}{\omega_0^2-\omega^2},
\end{equation}

\noindent where $\omega_0^2 = 1/\left(LC\right)$. Using the radiative correction factors, the final polarizability has the form

\begin{equation}
    \alpha_m(\omega) = \frac{\alpha_{m0} \omega_0^2}{\omega_0^2-\omega^2+j \alpha_{m0} \omega_0^2\left[ \frac{\beta}{a b}+\frac{k^3}{3 \pi}\right]},
\end{equation}

\noindent and

\begin{equation}
    \alpha_{m0} = \frac{a b L}{2 \mu}.
\end{equation}

\noindent Note that the form of the polarizability indicates that at low frequencies the polarizability approaches the static value found previously for a series inductance. 

The form of the resonance provides a useful means of extracting the effective inductance and capacitance parameters of a waveguide-fed metamaterial element. At frequencies well below the resonance, the polarizability is seen to converge to its static value, which is entirely inductive and can be determined from the above equation. The capacitance can then be directly found from the resonance frequency, or $C=1/\left(L\omega_0^2\right)$.

For metamaterial elements that may have some capacitive contribution, but operate at frequencies well below the resonance frequency, the polarizability shows some dispersion and can be well-approximated by 

\begin{equation}
    \alpha_m(\omega) \rightarrow \frac{\alpha_{m0}}{1-\frac{\omega^2}{\omega_0^2}}.
    \label{eq:approximate_alpha_for_slot}
\end{equation}

\noindent This form of the polarizability describes waveguide fed rectangular slots quite well, as will be shown below.

Also of interest is that, at resonance, the polarizability becomes entirely independent of the static polarizability value, reaching the peak value of

\begin{equation}
    \alpha_m(\omega_0) = \frac{-j}{\frac{\beta}{ab}+\frac{k^3}{3\pi}}.
    \label{eq:polarizability_at_resonance}
\end{equation}

\noindent It should be noted that (\ref{eq:polarizability_at_resonance}) indicates that regardless of the geometry of the metamaterial element, the peak polarizability has the \emph{same value}, determined solely by the radiation and waveguide damping terms. This result pertains only to iris-type elements where the waveguide thickness is much smaller than the wavelength, and not to compound elements such as the slot-fed patch considered below. In the absence of resistive losses (\ref{eq:radiated_power}) can be used with (\ref{eq:polarizability_at_resonance}) to find the fraction of incident power radiated into free space at resonance as

\begin{equation}
    \bar{P}_{rad}(\omega_0)=\frac{\left( \frac{k^3}{3\pi}\right)\left( \frac{2\beta}{ab}\right)}{\left( \frac{\beta}{ab}+\frac{k^3}{3\pi}\right)^2}.
\end{equation}

\noindent This last equation illustrates clearly that the peak power radiated is independent of the properties of the element and that there are no means of modifying or changing this coupling for a simple iris. It should be noted that while the peak polarizability is independent of $\alpha_{m0}$, the width of the resonance does depend on $\alpha_{m0}$; thus, the impact of having a larger static polarizability (or larger effective inductance) is to increase the width of the resonance (or decrease the effective quality factor).




\subsubsection{Aperture-Fed Elements and Similar}
\label{section:aperture_fed_elements}

The resonant slot-like metamaterial element has the disadvantage that its radiative coupling cannot be controlled and is generally quite large. For waveguides that support propagation along only one dimension, the large coupling suggests that a metamaterial antenna would have very low aperture efficiency, as most of the radiation may be lost within just a few elements. Other metamaterial designs can provide a means of independently controlling the resonance of the element and the coupling of the element to the waveguide. The slot-coupled patch is one such element and is described in this section. 

\begin{figure}[!b]
	\centering
	\includegraphics[width=1.85in]{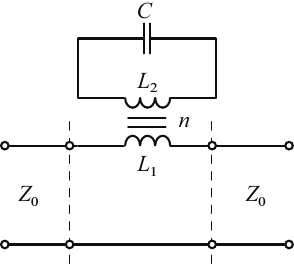}
	\caption{\label{fig:transformer_model}A simple model of a metamaterial element coupled to a resonator. As shown, resistive losses and radiative losses are neglected. The coupling between waveguide and resonator is modeled as an ideal transformer with turn ratio $n$. The simple resonator model is only appropriate for certain antenna types near their resonance.}
\end{figure}

Our goal here is to provide an intuitive understanding of the coupled element rather than deriving a rigorous model. A collection of equivalent circuit models for patch antenna variants can be found in the literature, based typically on a transmission line or a resonant cavity model. These models can be incorporated into the framework developed here. For the inital analysis, though, we consider a very simple resonator circuit model coupled to the waveguide via a transformer circuit, as shown in Fig. \ref{fig:transformer_model}.

Although the transformer model results are well-known, it is worth presenting a few of the key steps here. Treating both the waveguide feed as well as the region between the patch and upper conductor as transmission lines, we seek to determine the effective impedance as seen by the waveguide and to be used in the formulas derived above. Assuming the voltage across inductor $L_1$ is $V$, we have

\begin{equation}
\begin{aligned}
\label{eq:transformer_circuit_equations}
    V-j \omega L_1 i_1 - j \omega M i_2 &= 0, \\
    j \omega L_2 i_2 +j \omega M i_1 + \frac{1}{j \omega C}i_2 &= 0,
\end{aligned}
\end{equation}


\noindent where $M$ is the mutual inductance between the two inductors. The impedance, as seen from the waveguide side, is then

\begin{equation}
    Z = \frac{V}{i_1}=j \omega L_1 + j\omega M \frac{i_2}{i_1},
\end{equation}

\noindent From the second of the circuit equations, we can write

\begin{equation}
    \frac{i_2}{i_1} = \frac{-j \omega M}{j \omega L_2 + \frac{1}{j\omega C}}.
\end{equation}

\noindent The impedance can then be written as

\begin{equation}
    Z = j \omega L_1 +   \frac{\omega^2 M^2}{j\omega L_2 + \frac{1}{j\omega C}}.
\end{equation}

\noindent Assuming an ideal transformer, such that $M^2 = L_1 L_2$,

\begin{equation}
    Z = j \omega L_1 \left[1-\frac{j\omega L_2}{j\omega L_2 + \frac{1}{j\omega C}}\right],
\end{equation}

\noindent which can be in turn written as

\begin{equation}
Z = \frac{j\omega L_1 \omega_0^2}{\omega_0^2-\omega^2},
\end{equation}

\noindent where $\omega_0^2 = 1/(L_2 C)$.

In the final expression for the impedance, we see that the resonance frequency is set by the characteristics of the external resonator ($L_2$ and $C$), with the inductance of the metamaterial element (or iris) serving as a coupling between the waveguide and resonator. This final equation is nearly identical to that of the simple metamaterial resonator above, except that the coupling inductance is related to the resonator inductance by

\begin{equation}
    L_1 = \frac{L_2}{n^2}.
\end{equation}

\noindent We thus find the static polarizability for the iris-fed patch and similar geometries is

\begin{equation}
    \label{eq:resonant_polarizability_patch}
    \alpha_m = \frac{a b L_2 \omega_0^2}{2\mu n^2}\frac{1}{\omega_0^2-\omega^2},
\end{equation}

\noindent where $\omega_0^2 = 1/(L_2 C)$. Using the radiative correction factors, the final polarizability has the form

\begin{equation}
    \alpha_m(\omega) = \frac{\alpha_{m0} \omega_0^2}{\omega_0^2-\omega^2+j \alpha_{m0} \omega_0^2\left[ \frac{\beta}{a b}+F\frac{k^3}{3 \pi}\right]},
    \label{eq:polarizability_with_coupling}
\end{equation}

\noindent and

\begin{equation}
    \alpha_{m0} = \frac{a b L_2}{2 n^2 \mu}.
\end{equation}

\noindent In the radiation damping term, note that we have effectively removed the factor of $n^2$, since the free-space radiation term will relate to the effective magnetic moment external to the waveguide, which may also introduce a factor $F$. At resonance, then, the fraction of power radiated versus the incident power becomes

\begin{equation}
    \bar{P}_{rad}(\omega_0)=\frac{\left( \frac{F n^2 k^3}{3\pi}\right)\left( \frac{2\beta}{ab}\right)}{\left( \frac{\beta}{ab}+\frac{F n^2 k^3}{3\pi}\right)^2}.
    \label{eq:fraction_radiated_power_resonator}
\end{equation}

\noindent This last equation enables the transformer turn ratio to be extracted from simulations. As the coupling decreases, $n$ becomes smaller than unity and the radiation fraction is reduced as expected.

In the above analysis we have assumed the hypothetical antenna presented simply a capacitive load to the slot. A patch antenna, by contrast, must be modeled with a more realistic circuit. Here, we treat the patch antenna as a center-fed transmission line, with either end terminated in an open (magnetic) boundary condition. With this approximation the impedance of the line is $Z(\omega)=-j 2 Z_p \cot{\beta_p l/2}$, where the patch length is $l$, $\beta_p = \omega/c$ is the propagation constant within the patch and $Z_p=\eta d/w$ is the characteristic impedance of the patch. Replacing the $1/(j\omega C)$ term in (\ref{eq:transformer_circuit_equations}) with the total impedance $Z(\omega)$ yields

\begin{equation}
    Z = j \omega L_1 \frac{-j 2 Z_p \cot{\beta_p l/2}}{j \omega L_2 - j 2 Z_p \cot{\beta_p l/2}}.
\end{equation}

\noindent The static polarizability is then

\begin{equation}
    \alpha_{m} = \frac{ab L_2}{2 \mu n^2}\frac{2 \omega \frac{Z_p}{L_2} \cot{\beta_p l/2}}{-\omega^2 + 2 \omega \frac{Z_p}{L_2} \cot{\beta_p l/2}}
\end{equation}

\noindent Finally, adding in the radiation damping terms we obtain

\begin{equation}
    \alpha_m(\omega) =   \frac{\alpha_{m0}u}{u-\omega^2 +j \alpha_{m0} u \Big[ \frac{\beta}{ab}+F n^2 \frac{k^3}{3 \pi}\Big]}
    \label{eq:patch_polarizability}
\end{equation}

\noindent with

\begin{equation}
    u = 2 \omega \frac{Z_p}{L_2}\cot{\beta_p l /2}
\end{equation}

\noindent Note that the resonance of the bare patch occurs at $\beta_p l/2 = \pi/2$, or when $l=\lambda_p/2$. The resonance frequency of the coupled patch is shifted, however, and can be found from

\begin{equation}
    2\omega_0 \frac{Z_p}{L_2}\cot{Z_p C_l l \omega_0}=\omega_0^2.
\end{equation}

The model presented here is admittedly crude and meant to provide some insight into resonant elements whose coupling can be modified by design. An improved model would take into account the fringing capacitance at the patch edges as a shunt admittance, and would properly determine the factor $F$ to be used in the radiation damping term. We do not pursue these models further, but rather use the above equation to deduce some of these factors and understand the line shapes that are found from full-wave simulations. 

\subsubsection{Magnetic Polarizability with Resistive Losses}

In the case that there are resistive losses in addition to the radiation damping terms, we can introduce a resistance in series with the inductance by letting 

\begin{equation}
    L \rightarrow L+\frac{R}{j \omega}.
\end{equation}

\noindent Inserting the above into (\ref{eq:resonant_polarizability}) we obtain for the static polarizability

\begin{equation}
    \alpha_m = \alpha_0 \omega_0^2 \Big( \frac{1-j\frac{\Gamma}{\omega}}{\omega_0^2 - \omega^2 +j \omega \Gamma}\Big),
\end{equation}

\noindent where $\Gamma = R/L$ is the damping factor due to resistive losses. Using the static polarizability with the radiation corrections yields

\begin{equation}
    \alpha_m = \frac{\alpha_0 \omega_0^2\left(1-j\Gamma/\omega\right)}{\omega_0^2-\omega^2 +j \omega \Gamma + j\alpha_0 \omega_0^2\left(1-j\Gamma/\omega\right)\left(\frac{\beta}{ab}+\frac{k^3}{3\pi}\right)}.
\end{equation}

\noindent At frequencies well below the resonance, the polarizability is complex, and we find

\begin{equation}
    \textrm{Im}\{\alpha_m\}=-\alpha_0 \frac{\Gamma}{\omega} = -\frac{abR}{2\mu \omega}.
\end{equation}

\noindent At low frequencies, then, the effective resistance can be determined from the extracted polarizability using the above formula, allowing resistive losses to be differentiated from radiative losses. 

Closer to the resonance, we can neglect terms with $\Gamma/\omega$, so that

\begin{equation}
    \alpha_m = \frac{\alpha_0 \omega_0^2}{\omega_0^2-\omega^2 +j \omega \Gamma + j\alpha_0 \omega_0^2 \left(\frac{\beta}{ab}+\frac{k^3}{3\pi}\right)}.
\end{equation}

\noindent Here, we see that the resistive damping simply adds the radiative damping terms leading to a broader (lower Q) resonance and the loss mechanisms are not distinguishable. Extrapolating the resonance curve to low frequency and finding the static polarizability $\alpha_0$ provides the means of determining all of the effective circuit parameters.

\subsection{ABCD Matrix: Electric Dipole}\index{ABCD Matrix: Electric Dipole}

Using the relationship of the S-parameters to the \emph{ABCD} matrix, we can perform the same analysis assuming only an electric dipole. In this case, we obtain $T_{11}=T_{22}=1$, $T_{12}=0$, and

\begin{equation}
    T_{21}=\frac{1}{Z_0} \frac{2 j A_e/2}{1-j A_e/2} = Y,
\end{equation}

\noindent Following the same steps as with the magnetic polarizability, it is straightforward to show that

\begin{equation}
    \textrm{Im}\{ A_e \} = -\frac{|A_e|^2}{2},
\end{equation}

\noindent and thus

\begin{equation}
    Y = j Y_0 \frac{|A_e|^2}{\textrm{Re}\{A_e \}},
\end{equation}

\noindent from which we can determine the electric polarizability as

\begin{equation}
    \alpha_{e0} = -j \frac{\beta a b}{k^2}\frac{Y}{Y_0},    
\end{equation}

\noindent a result also consistent with \cite{collin1990fieldCh7}.

\begin{figure}[!b]
	\centering
  \includegraphics[width=2.05in]{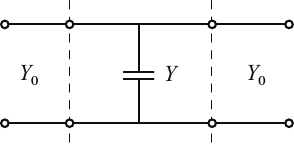}
    \caption{Equivalent circuit for the case of an electric polarizability.}
    \label{fig:shunt_admittance_network}
\end{figure}

In the absence of radiative damping, the electric polarizability equats to a shunt capacitance, as illustrated in Fig. \ref{fig:shunt_admittance_network}.

Although the electric dipole is generally of much less importance for the types of metasurface antennas of interest here, we provide the equivalent electric polarizability after following the same procedure that led to (\ref{eq:magnetic_polarizability_circuit_equiv}). We find

\begin{equation}
    \alpha_e = \frac{\frac{ab}{2 \epsilon}C}{1+j\frac{k^2}{\beta a b}\alpha_{e0}}
\end{equation}

\noindent where $C$ is the shunt capacitance. For low frequencies we find the static electric polarizability to be

\begin{equation}
    \alpha_{e0} = \frac{ab}{2\epsilon}C
\end{equation}

\subsection{ABCD Matrix: Magnetic and Electric Dipole}\index{ABCD Matrix: Magnetic and Electric Dipole}

\begin{figure}[!b]
	\centering
  \includegraphics[width=2.05in]{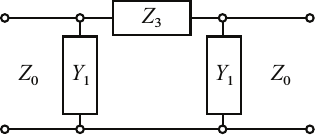}
    \caption{Equivalent circuit for the case of both an electric and a magnetic polarizability.}
    \label{fig:pi_network}
\end{figure}

In the case that both dipole components exist and are significant, then the \emph{ABCD} matrix can be computed as above with both electric and magnetic polarizability terms. However, to facilitate the analysis, we anticipate that the equivalent circuit might conform to either a traditional $\pi$ or \emph{T} network. Here we consider the $\pi$ network, as shown in Fig. \ref{fig:pi_network}, which consists of a series impedance $Z_3$ in between two shunt admittances $Y_1$. The \emph{ABCD} matrix elements can be computed from

\begin{equation}
\begin{aligned}
    \bar{\bar{T}} &= \begin{pmatrix}
    1 & 0  \\
    Y_1 & 1  
    \end{pmatrix}
    \begin{pmatrix}
    1 & Z_3  \\
    0 & 1  
    \end{pmatrix}
    \begin{pmatrix}
    1 & 0  \\
    Y_1 & 1  
    \end{pmatrix} \\ &=
    \begin{pmatrix}
    1+Z_3 Y_1 & Z_3  \\
    2 Y_1+Y_1^2 Z_3 & 1+Z_3 Y_1  
    \end{pmatrix}.
\end{aligned}
\end{equation}


\noindent Given the form of the transfer matrix, it is easiest to start by computing $S_{12}$, which will directly result in the impedance $Z_3$ being determined:

\begin{equation}
    T_{12}=j Z_0 \frac{\left(2j+A_e\right)A_m}{2j+A_e+A_m}=Z_3,
\end{equation}

\noindent from which we have

\begin{equation}
    Y_3 = \frac{1}{Z_3}=\frac{1}{2Z_0 j}\left[ \frac{1}{A_m/2}+\frac{1}{(j+A_e/2)}\right],
\end{equation}

\noindent or

\begin{equation}
\begin{aligned}
    Y_3 &= \frac{1}{2Z_0}\left[ \frac{1-j A_m/2}{j A_m/2}+\frac{jA_e/2}{(1-jA_e/2)}\right].
\end{aligned}
\end{equation}

\noindent We thus find

\begin{equation}
    Y_3 = \frac{1}{Z}+\frac{Y}{4},
\end{equation}

\noindent where $Z$ and $Y$ are the impedance of the magnetic polarizability and the admittance of the electric polarizability, respectively, when either are alone.
\noindent Using this result, we obtain for $T_{11}$ and $T_{22}$,

\begin{equation}
\begin{aligned}
    T_{11} &= 1+Y_{1} Z_{3}.
\end{aligned}
\end{equation}

\noindent From this last equation, we find that

\begin{equation}
    Y_1 = \frac{Y}{2},
\end{equation}

\noindent so that all of the elements of the $\pi$ equivalent circuit have been found. A similar network model is presented in \cite{collin1990fieldCh7} for two waveguides coupled by a small aperture---in that case presented as a "T" network---that shows a similar distribution of the shunt and series components.

\section{Illustrative Examples}  \label{sec:numerical examples}

We illustrate the application of the circuit model developed in Section \ref{sec:circuit model} using the scattering parameters obtained from full-wave simulations on selected metasurface element models. We focus here on metamaterial elements that exhibit a predominantly magnetic dipolar response, as they are simpler to analyze and are of relevance for metasurface antenna designs \cite{pulido2016discrete,pulido2017discrete}. The goal here is to show that the equivalent circuit model of the metamaterial element can be used in conjunction with added lumped components, allowing rapid prediction of the properties of the composite element. This capability becomes of great utility when designing active or tunable elements that must interface with additional electronic circuits.


\subsection{Rectangular Slot}

\begin{figure}[!b]
\centering
\includegraphics[width=3.25in]{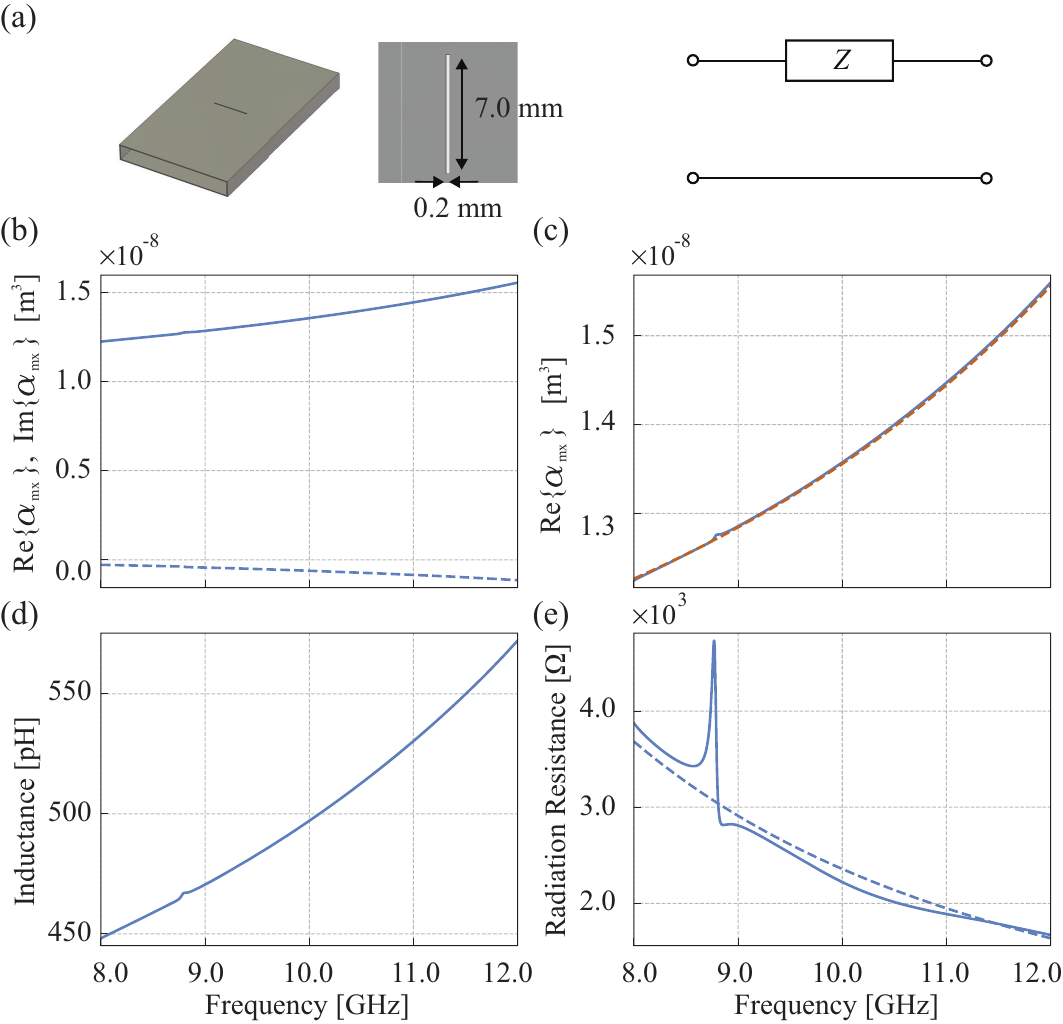}
\caption{(a) The slot waveguide and equivalent circuit model. (b) The real (solid curve) and imaginary (dashed curve) parts of the polarizability extracted from the computed S-parameters. (c) The computed (solid curve) and fitted (dashed curve) real part of the polarizability. (d) The effective inductance of the slot, as computed using (\ref{eq:static_polarizability}). (e) The computed (solid curve) radiation resistance, compared with (\ref{eq:radiation_resistance_formula}) (dashed curve).}
\label{fig:rectangular_slot_simulation}
\end{figure}

We here consider a simple rectangular slot inserted into the top wall of a rectangular waveguide, as shown in Fig. \ref{fig:rectangular_slot_simulation}(a). It is well-known that the rectangular slot radiates as a magnetic dipole, and thus should correspond well to the analysis of the non-resonant magnetic dipole presented in \ref{ABCD Matrix: Magnetic Dipole}. Here, the air-filled waveguide has a width $a=22.9$ mm and height $b=3$ mm---values chosen such that the waveguide operates at X-band frequencies (8-12 GHz), but with no other specific device or antenna in mind. The length and width of the slot are $7.0$ mm and $0.2$ mm, respectively, with a waveguide thickness of $0.1$ mm. The axial ratio of the slot was chosen to be such that the slot scatters the field predominantly as a magnetic dipole \cite{oliner1957impedance,balanis2005antenna,pulido2016discrete,pulido2017discrete}.

The rectangular slot is oriented such that the long axis is perpendicular to the direction of propagation of the feed wave, such that the element should behave as a series inductance. Were this to be true, then the polarizability would be described simply by (\ref{eq:static_polarizability}) and would be non-dispersive. To ascertain the behavior of an actual slot, a full-wave simulation was performed using CST Microwave Studio to obtain the S-parameters for the structure shown in Fig. \ref{fig:rectangular_slot_simulation}(a). The S-parameters were de-embedded to the center of the slot from either port. The length of the waveguide section was chosen (arbitrarily) to be $40$ cm, with the slot positioned directly in the center. The S-parameters were exported to a text file, and then imported into a custom Python script to compute the effective polarizability using (\ref{eq:polarizabilities_vs_s_parameters}).

As expected, the resulting polarizability is predominantly magnetic, with a value roughly of $\alpha_m \approx 1.0 \times 10^{-8} \mathrm{m^3}$ and plotted in Fig. \ref{fig:rectangular_slot_simulation}(b) ($\alpha_{ey}$, not shown, is roughly one hundred times smaller). The dispersion present in the polarizability suggests a non-negligible series capacitance is present, so that the appropriate model to describe the slot is that of (\ref{eq:resonant_polarizability}) with the approximation of (\ref{eq:approximate_alpha_for_slot}) applied. We perform a curve fit of the real part of the extracted $\alpha_m$ to (\ref{eq:approximate_alpha_for_slot}), from which we determine values of $\alpha_{m0}$ and $\omega_0$. From $\alpha_{m0}$ we can find the equivalent inductance $L$, with the equivalent capacitance found using $C = 1/(\omega_0^2 L)$. The curve fit is shown as the dashed line in Fig. \ref{fig:rectangular_slot_simulation}(c), with the relevant extracted parameters shown in Table \ref{table:1}. The effective inductance as a function of frequency is plotted in Fig. \ref{fig:rectangular_slot_simulation}.

\renewcommand{\arraystretch}{1.2}
\begin{table}
\caption{Parameters for the slot element}
\small
\centering
\begin{tabular}{ |p{2cm}|p{2cm}|p{2cm}|  }
\hline
\multicolumn{3}{|c|}{Fit Parameters}\\
\hline
Parameter & Value & Units \\
\hline
$\alpha_{m0}$ & 1.05 x$10^{-8}$ & $m^3$ \\
$\alpha_{e0}$ & 0.01 x$10^{-8}$ & $m^3$ \\
$f_0$ & 21.02 & GHz \\
$L_{eff}$ & 383.7 & pH \\
$C_{eff}$ & 0.15 & pF \\
\hline
\end{tabular}
\vspace*{2mm}
\label{table:1}
\end{table}

The parameters determined from the polarizability extraction show the approximate resemblance of the slot element to a series inductance, with the electric polarizability being negligible in comparison and with a relatively small series capacitance in parallel.

As an additional confirmation of the analytical predictions presented here, we also plot the radiation resistance as directly computed from the simulation versus that calculated in (\ref{eq:radiation_resistance_formula}). The curve for the simulated structure was obtained by calculating the series admittance $Y$ from the S-parameters and taking the real part of $1/Y$. Aside from an anomalous spike just below 9 GHz, the theory and simulation are in excellent agreement.

\subsection{Resonant Slot with a Capacitor}

Resonant metamaterial elements can be tuned with the inclusion of elements such as diodes or varactors, enabling some control over phase and amplitude of the scattered fields. With the equivalency of the rectangular slot to a series inductance confirmed, it is reasonable to consider adding a capacitive component to the element to introduce a resonance at the targeted frequencies. A lumped capacitor could be used to provide a fixed choice of phase for the element, while a varactor integrated into the circuit could provide a voltage-controlled phase shift. The details of integrating such devices into the metamaterial element necessitate the discussion of DC bias circuits, which is beyond the scope of the analysis presented here; instead, our focus is on the simple inclusion of lumped components with the metamaterial element. 

Lumped components can be introduced into the simulation domain in most of the standard commercial electromagnetic solvers. In CST Microwave Studio, used here, one specifies values for the inductance, capacitance and resistance of the lumped component, as well as two conducting points that will be spanned by the element. Before proceeding, however, it is important to assess the impact of this operation. The effective leads of the lumped component may introduce an additional inductance beyond that contained in the actual lumped component, which will need to be accounted for in the final circuit. This additional inductance we refer to here as the package inductance.

To assess the package inductance of the lumped component, we first set the capacitance of the element to something very large (such as 50 pF), which essentially makes the lumped component an effective short, aside from any package inductance that might exist. With the same extraction procedure followed as described above, the parameters associated with the composite element are shown in Table \ref{table:2}. We see here that the effective inductance is significantly smaller, indicating that the addition of the lumped component indeed added a significant amount of inductance to the circuit. Assuming the two inductances are in parallel, we can derive a value for the package inductance of the lumped element as $L_p=360$ pH.

\renewcommand{\arraystretch}{1.2}
\begin{table}
\caption{Parameters for the shorted slot element}
\small
\centering
\begin{tabular}{ |p{2cm}|p{2cm}|p{2cm}|  }
\hline
\multicolumn{3}{|c|}{Fit Parameters}\\
\hline
Parameter & Value & Units \\
\hline
$\alpha_{m0}$ & 0.51 x$10^{-8}$ & $m^3$ \\
$\alpha_{e0}$ & 0.01 x$10^{-8}$ & $m^3$ \\
$f_0$ & 30.85 & GHz \\
$L_{tot}$ & 185.72 & pH \\
$L_{p}$ & 360.72 & pH \\
$C_{eff}$ & 0.14 & pF \\
\hline
\end{tabular}
\vspace*{2mm}
\label{table:2}
\end{table}

\begin{figure}[!b]
\centering
\includegraphics[width=3.25in]{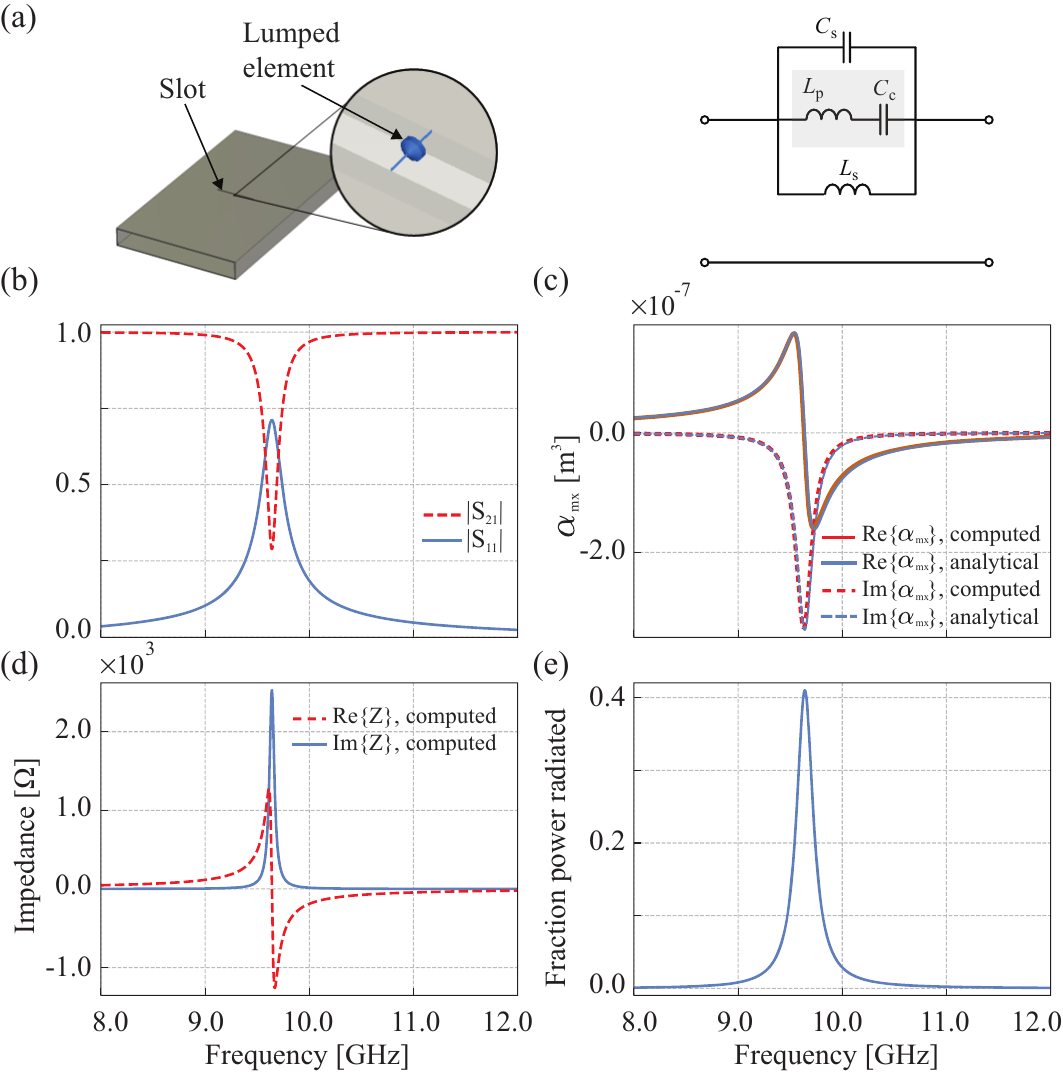}
\caption{(a) The capacitor loaded-slot structure and equivalent circuit model. (b) The computed magnitudes of the S-parameters. (c) The computed (blue) and analytically determined (red) polarizability, real parts correspond to the solid curves while imaginary parts correspond to the dashed curves. (d) The computed series impedance. (e) The fraction of power radiated, computed from the extracted polarizability using (\ref{eq:radiated_power}).}
\label{fig:rectangular_slot_with_cap_simulation}
\end{figure}

With the values of the slot inductance $L_s$ and capacitance $L_c$ now determined, as well as the effective package inductance $L_p$ of the lumped element, we can now assess the effective polarizability of a slot with a lumped capacitance $C_c$. The equivalent circuit model is shown in Fig. \ref{fig:rectangular_slot_with_cap_simulation}(a). This more complicated series impedance must be used in place of the simple inductance found previously. The impedance can be evaluated as

\begin{equation}
    Z^{-1} = \frac{1}{j \omega L_s} + j \omega C_s + \frac{1}{j \omega L_p + \frac{1}{j \omega C_c}},
\end{equation}

\noindent which can be written as

\begin{equation}
    Z = \frac{j \omega}{C_s} \frac{\omega_c^2-\omega^2}{(\omega_s^2-\omega^2)(\omega_c^2-\omega^2)-\omega^2 \omega_m^2},
\end{equation}

\noindent with $\omega_s^2 = 1/(L_s C_s)$, $\omega_c^2 = 1/(L_p C_c)$, and $\omega_m^2 = 1/(L_p C_s)$. The resonance frequencies of the element occur at the poles of the above equation, which we write as $\omega_0^+$ and $\omega_0^-$. The lowest resonance frequency is the one that is of interest here, so that the effective static polarizability can be written, using (\ref{eq:generalized_polarizability}),

\begin{equation}
    \alpha_{m} = \frac{a b}{2 \mu C_s}  \frac{\omega_p^2-\omega^2}{(\omega_0^+)^2-\omega^2}\frac{1}{(\omega_0^-)^2-\omega^2}
\end{equation}

\noindent Applying the radiative corrrections, as done above, we obtain

\begin{equation}
    \alpha_m(\omega) = \frac{\alpha_0 (\omega_p^2-\omega^2)}{(\omega_0^-)^2-\omega^2+j \alpha_0 (\omega_p^2-\omega^2) \left[ \frac{\beta}{a b}+\frac{k^3}{3 \pi}\right]},
    \label{eq:slot_with_cap_polarizability}
\end{equation}

\noindent from which we can determine the static polarizability as

\begin{equation}
    \alpha_0 = \frac{a b}{2 \mu C_s}  \frac{1}{(\omega_0^+)^2-\omega^2}
\end{equation}

\noindent Because the second resonance frequency will be typically much higher than the frequencies of interest, a reasonable approximation to the above is

\begin{equation}
    \alpha_0 = \frac{a b}{2 \mu C_s}  \frac{1}{(\omega_0^+)^2}
\end{equation}

\noindent Within the same assumption, the upper resonance frequency can further be approximated as

\begin{equation}
    (\omega_0^+)^2 \approx \frac{1}{L_s C_s} + \frac{1}{L_p C_s}
\end{equation}

\noindent from which we obtain an approximation to the static polarizability as

\begin{equation}
    \alpha_0 \approx \frac{a b}{2 \mu}  \frac{L_s L_p}{(L_s + L_p)}
\end{equation}

\noindent While this final expression is satisfying in its interpretation as relating the static polarizability to the total inductance of the element---and, indeed, the expression is valid for the example presented here---nevertheless it should be used with caution since the underlying assumptions may not always be valid. The generalized impedance, however, is always valid and can be used to obtain the effective polarizability.

One final note is necessary regarding the use of a lumped capacitor with the slot. In determining the effective impedance of the equivalent transmission line circuit model, we make use of an effective voltage, current and line impedance that are related to the average power in the waveguide mode \cite{pozar2009microwave}. The lumped element is placed across the slot, however, at a specific position in the waveguide (here, the midplane) where the voltage is specifically defined. We must relate this voltage to the effective transmission line voltage. For our particular case, we can expect that this voltage will be about a factor of $\sqrt{2}$ greater than the line voltage, while the transmission line current is continuous through the element. The transmission line impedance, therefore is reduced by a factor of $\sqrt{2}$, and thus the effective capacitance can be understood as being increased by a factor of $\sqrt{2}$. This factor provides the only uncertainty in the model, and may require slightly adjusting the effective $C_c$ to reflect the correct local voltage.

With the model set forth as above and using the values of $L_s$, $L_p$ and $C_s$ determined from the first two simulations, we select a value for $C_c$ that will place a resonance within the X-band region. Here, we select a value for the element of $C_c = 0.2$ pF. The magnitudes of the S-parameters are shown in Fig. \ref{fig:rectangular_slot_with_cap_simulation}(b), indicating the level of coupling between the waveguide and the element. The extracted polarizability is shown in Fig \ref{fig:rectangular_slot_with_cap_simulation}(c). Here, we have also plotted (\ref{eq:slot_with_cap_polarizability}) for comparison using the values given above, with the factor of $\sqrt{2}$ included for $C_c$. The agreement between the simulated structure and analytical prediction is exact; the curves lie on top of each other. 

It should be noted that the expression for the polarizability has no free parameters, providing a specific prediction for the resonance curve corresponding to the capacitor-loaded slot. The peak of the resonance, as discussed earlier, is determined entirely by the radiative damping factors, while the width of the curve is determined by a combination of the radiation damping parameters and effective polarizability. Likewise, the location of the resonance is determined by the poles of the impedance, so that all aspects of the effective polarizability are determined from the extracted $L_s$, $L_p$, $C_s$ and the lumped $C_s$ introduced.

The series impedance of the capacitor-loaded slot is shown in Fig. \ref{fig:rectangular_slot_with_cap_simulation}(d). The real part of the impedance corresponds to the radiation resistance, which can be used to determine the power radiated. Alternatively, (\ref{eq:radiated_power}) can be used with the computed S-parameters to determine the fraction of incident power radiated, as shown in Fig. \ref{fig:rectangular_slot_with_cap_simulation}(e). The peak of the radiated power can be predicted from (\ref{eq:fraction_radiated_power_resonator}), which yields a value of $\bar{P}_{rad}=0.41$, in close agreement with the value found from the simulated curve of Fig. \ref{fig:rectangular_slot_with_cap_simulation}(e). These comparisons provide a strong validation of the model.

\subsection{Slot-Coupled Patch Antenna}

\begin{figure}[!b]
\centering
\includegraphics[width=3.25in]{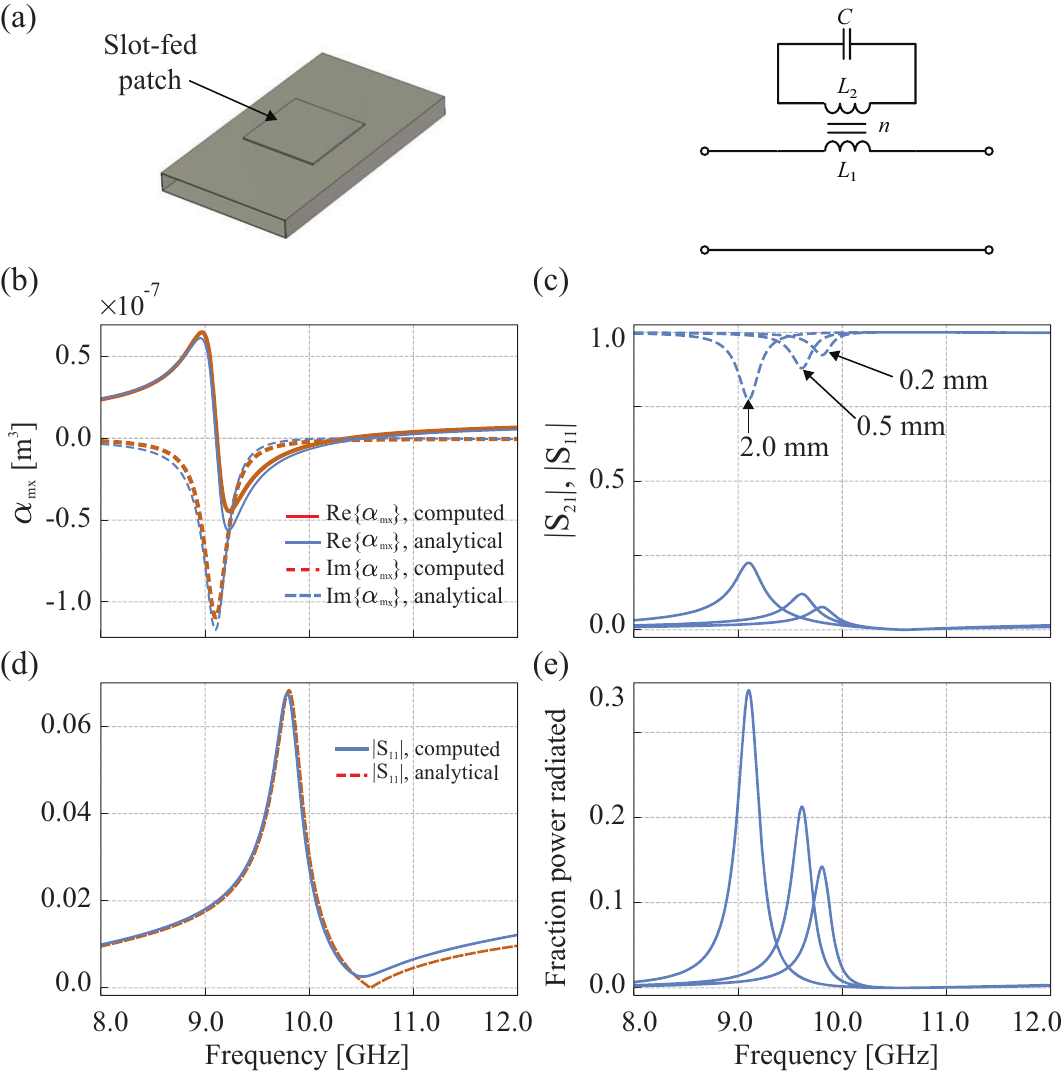}
\caption{(a) The slot-coupled patch antenna and an equivalent circuit model. Here, a transmission line impedance replaces the single capacitor shown, as indicated by (\ref{eq:patch_polarizability}). (b) The computed (blue) and analytically determined (red) real and imaginary parts of the polarizabilities for the $w=2.0$ mm slot. (c) The computed magnitudes of $|S_{21}|$ (dashed curves) and $S_{11}$ (solid curves) for slot widths of $0.2$, $0.5$ and $2.0$ mm. (d) The computed (blue) and analytically determined (red) $|S_{11}|$ curves for the $0.2$ mm patch. (e) The fraction of power radiated for patches with slot widths of $0.2$, $0.5$ and $2.0$ mm, computed from the extracted polarizability using (\ref{eq:radiated_power}).}
\label{fig:alpha_for_patch_figure}
\end{figure}

As a final example, we consider here the polarizability of a slot-coupled patch antenna to illustrate how the coupling to the waveguide mode can be controlled by design. For this example we use the same waveguide dimensions as in the previous sections, with a slot that is $l=6$ mm in length and with widths $w=0.2$ mm, $w=0.5$ mm, and $w=2.0$ mm. The patch is a perfectly conducting square with dimensions $13.0$ mm $\times$ $13.0$ mm, a thickness of $0.25$ mm, and placed a distance of $0.4$ mm above the surface of the waveguide and symmetric about the slot (Fig. \ref{fig:alpha_for_patch_figure}(a)). The slot-fed patch is an extremely common antenna configuration that has been amply studied \cite{pozar1986reciprocity,sullivan1986analysis,gauthier199994}; our goal here is to show how this antenna paradigm can be included within the polarizability description. While the polarizability extraction can be generally performed on a structure such as this, it is useful to understand the underlying mechanisms at play from an analytical perspective.


\renewcommand{\arraystretch}{1.2}
\begin{table}
\caption{Parameters Extracted for Slot-Coupled Patches}
\small
\centering
\begin{tabular}{ |p{1.5cm}|p{1.cm}|p{1.cm}|p{1.cm}|p{1.cm}|  }
\hline
\multicolumn{5}{|c|}{Slot Width (mm)}\\
\hline
Parameter & $0.2$ & $0.5$ & $1.0$  & $2.0$ \\
\hline
$f_0$ (GHz) & $9.8$ & $9.6$ & $9.3$ & $9.1$ \\
$n$ & $0.44$ & $0.42$ & $0.43$ & $0.42$ \\
$F n^2$ & 29 & 18 & 13 & 9 \\
$L_{1}$ (pH) & 201 & 283 & 335 & 464 \\
$L_{2}$ (pH) & 39 & 50 & 65 & 82 \\
$\bar{P}_{rad}$ & 0.12 & 0.19 & 0.25 & 0.32 \\
\hline
\end{tabular}
\vspace*{2mm}
\label{table:3}
\end{table}

We apply the theory presented in Section \ref{section:aperture_fed_elements} to model the slot-coupled patch. As discussed in that section, the model for the patch we apply is a first pass attempt, neglecting the capacitive load at the antenna edges and other effects in favor of a simpler description. As a first example, we perform a simulation for the case of the slot width of $w=2.0$ mm, comparing the extracted polarizability with (\ref{eq:patch_polarizability}). The result, shown in Fig. \ref{fig:alpha_for_patch_figure}(b), reveals acceptable but not precise agreement between the two curves. The fitting was performed by hand, adjusting the values of $L_2$, $n$ and $F$ until agreement was found. The general shape of the curve predicted by (\ref{eq:patch_polarizability}) precludes a perfect match with the extracted polarizability, but this is expected due to the incomplete model. Nevertheless, the general features of the polarizability curves are present and allow the key parameters to be extracted. Table \ref{table:3} shows the resonance frequency for this case is $9.1$ GHz, which implies a value of $L_2=82$ pH. The resonance frequency of the uncoupled patch is at $10.4$ GHz, so that the presence of the slot and its inherent inductance contribute to the frequency shift. Once the frequency of the resonance is found by adjusting $L_2$, the real and imaginary parts of the polarizability are matched by next adjusting $F$ and $n$. A value of $n$ is found via this process as $0.42$. This value can be tested against the expected value, which is the length of the slot divided by the width of the patch, or $6/13.6=0.44$. The agreement here is excellent and is very close to the value found in all four of the simulated structures considered here (see Table \ref{table:3}).

Given that the coupling element and patch alter the balance between scattering within the waveguide and to free space, it can be expected that the overall impact of the load seen on the waveguide side will vary with the slot dimensions. That this is the case can be seen in Fig. \ref{fig:alpha_for_patch_figure}(c), which shows the computed $|S_{21}|$ (dashed) and $|S_{11}|$ (solid) curves for the $w=0.2$, $w=0.5$ and $w=2.0$ mm patches. As the slot width increases, the coupling becomes stronger and also shifts the resonance to lower frequencies. From the analytical model, the shift to lower frequencies occurs due to the increase in the effective inductance of the slot. Also, as the slot width becomes larger, the factor $F n^2$ decreases, resulting in an increased coupling of the element to the waveguide; this last result is intuitive, as it would be expected the larger slot would present a greater coupling to the patch resonator and would therefore be a larger perturbation within the waveguide. 

While the analytical model is not perfect, it yet captures many of the key features of the slot-fed patch, including a kink that occurs at the frequency where the uncoupled patch would have its resonance. From (\ref{eq:patch_polarizability}) it can be seen that the polarizability will have a zero at this frequency. In Fig. \ref{fig:alpha_for_patch_figure}(d) we plot the absolute value of $S_{11}$ for the computed and analytical cases of the $w=0.2$ mm patch. Although the zero is more pronounced for the analytical case, the overall curves are in rough agreement. The shape of the curves deviate considerably from a simply resonator model, suggesting the simplified transmission line model is a useful initial description of the mechanisms at play in the slot-coupled patch.

Coincident with the varying coupling of the patch to the waveguide is a variation in the fraction of incident power radiated to free space. Indeed, the fraction of power radiated at resonance is found to increase with increasing coupling, starting from a value of $\bar{S}_{rad}(f_0)=0.12$ for the $w=0.2$ mm patch, to a value of $\bar{P}_{rad}(f_0)=0.32$ for the $w=2.0$ mm patch. The results (computed from the simulations) are shown in Fig. \ref{fig:alpha_for_patch_figure}(e).

\section{Conclusion}
Our goal in this work has been to present a reconciliation of the polarizability description of waveguide-fed metamaterial elements with more traditional, antenna engineering-oriented approaches. The connection between the polarizability model and traditional models is enabled by consideration of the computed S-parameters, which can be related directly to equivalent circuit models that describe the dipole elements in terms of more familiar circuit quantities. As a result of this analysis, the role of radiation damping proved to be of significance in understanding the analytical forms of the equivalent circuit models found. For example, in the dipole description it is not possible to describe an element as having a purely real polarizability, since scattering within the waveguide must be accounted for explicitly through an imaginary component. By contrast, an element can be described as purely reactive within the circuit model (no resistance term necessary). This difference leads to the particular analytical forms found for the polarizability expressions, from which we are able to extricate the correct circuit equivalents.

Once the circuit parameters have been found, the question naturally arises as to whether these values can be considered \emph{actual}---that is, whether a metamaterial element described as an inductance and capacitance would behave as modeled in the presence of additional lumped circuit elements. The analysis and simulation studies presented here confirm that this is indeed the case; once the appropriate circuit model has been identified, additional lumped elements can be integrated into the element and the properties of the composite described with an equivalent polarizability. This capability is of great importance for functional metamaterial elements that enable tuning or switching via the inclusion of lumped circuit elements. While full-wave simulations can generally be used to compute the behavior of such elements, the analytical models provide greater insight and can further be used to circumvent additional full-wave simulations for extremely rapid modeling of largescale apertures.

In the present work we have considered just a few metamaterial elements to demonstrate the foundations of the approaches developed. More complicated metamaterial elements, such as the commonly-used cELC, can be incorporated within the framework presented here via a suitable equivalent circuit model. It should be noted that the basic cELC can be described extremely well by the simple resonator model presented above, with only a parallel capacitance and inductance used; however, such a model will not be valid if lumped elements are to be integrated into the element. For such models, a collection of capacitances and inductances describing specific slots and sections of the metamaterial element would be the desirable description---a task left for future work.

The dipole framework for metasurface design is motivated for apertures comprising elements, each smaller than than the wavelengths of interest, and excited by a common feed, such as a waveguide or free-space illuminating beam. Such elements are conveniently and efficiently described as polarizable dipoles, with the entire aperture treated as a collection of interacting dipoles. Although the polarizability description has proven useful in this regard, the notion of a polarizability and its intuitive meaning can be confusing in many contexts, leading to an opaque design process. By establishing a rigorous connection of these more physics-based definitions to more familiar circuit-based definitions, we provide an alternative approach that allows for even more complex elements to be considered.

\appendix[Polarizability Extraction Equations]

Consider a waveguide mode incident on an infinitesimally thin sheet inserted into the guide and having both an electric and a magnetic conductivity, located at $z=0$. The boundary conditions with respect to the sheet can be determined from Maxwell's equations:

\begin{equation}
\begin{aligned}
    \nabla \times \vec{E} &= -\mu \frac{\partial \vec{H}}{\partial t} - \vec{J}_m, \\
    \nabla \times \vec{H} &= \epsilon \frac{\partial \vec{E}}{\partial t} + \vec{J}_e,         
\end{aligned}
\end{equation}

\noindent where

\begin{equation}
\begin{aligned}
    \vec{J}_e &= \frac{\partial \vec{P}}{\partial t} = j \omega \vec{P}, \\
    \vec{J}_m &= \mu \frac{\partial \vec{M}}{\partial t} = j \mu \omega \vec{M}.
\end{aligned}
\end{equation}

We designate region 1 as the region for which $z<0$ and region 2 as the region for which $z>0$, with the wave incident from $z=-\infty$. If we integrate the above equations over a loop that extends between $z=-\Delta/2$ to $z=\Delta/2$, where $\Delta$ is small enough that the time derivative terms can be neglected, then:

\begin{equation}
\begin{aligned}
    \oint \vec{E} \cdot d\vec{l} &= -j \omega \mu \int \vec{M} \cdot d\vec{S}, \\
    \oint \vec{H} \cdot d\vec{l} &= -j \omega \int \vec{P} \cdot d\vec{S}, \\    
\end{aligned}
\end{equation}

\noindent from which we arrive at the boundary conditions

\begin{equation}
\begin{aligned}
    E_{1,y} - E_{2,y} &= j\omega \mu M_x \Delta, \\
    H_{1,x} - H_{2,x} &= j\omega P_y \Delta.
\end{aligned}    
\end{equation}

We next assume that the polarization $\vec{P}$ and magnetization $\vec{M}$ relate to the incident field by a polarizability, or 

\begin{equation}
\begin{aligned}
    P_y &= \frac{\epsilon \alpha_{ey}}{ab\Delta}E_{inc}, \\
    M_x &= \frac{\alpha_m}{a b \Delta}H_{inc} = \frac{\alpha_{mx}}{Z_0 a b \Delta}E_{inc}.
\end{aligned}
\end{equation}

By definition, the incident, transmitted and reflected fields can be defined as

\begin{equation}
\begin{aligned}
    E_{1,y} &= (E_{inc}+r E_{inc}) = E_{inc} (1+S_{11}), \\
    E_{2,y} &= E_{inc} t  = E_{inc} S_{21}, \\
    H_{1,x} &= \frac{E_{inc}}{Z_0}(1-r) = \frac{ E_{inc}}{Z_0} (1 - S_{11}), \\
    H_{2,x} &= t\frac{E_{inc}}{Z_0} = \frac{  E_{inc}}{Z_0}S_{21}.
\end{aligned}
\end{equation}

\noindent Using these forms with the boundary conditions we obtain

\begin{equation}
\begin{aligned}
    1+S_{11}-S_{21} &= j \frac{\mu \omega \alpha_{mx}}{Z_0 a b}=j \frac{\beta}{ab}\alpha_{mx}, \\
    1-S_{11}-S_{21} &= j \frac{Z_0 \omega \epsilon \alpha_{ey}}{a b}=j \frac{k^2}{a b \beta}\alpha_{ey},
\end{aligned}
\end{equation}

\noindent where we have used $Z_0 = \mu \omega/\beta$.


%



\section*{Acknowledgment}

The authors are grateful to Andreas Barchanski (Dassault Systemes and TU Darmstadt) for confirming the manner in which CST Microwave Studio treats lumped elements. This work was partially funded by Kymeta Corporation.

\bibliographystyle{IEEEtran}
\bibliography{bibtex/bib/IEEEexample}

\end{document}